\documentclass[fleqn,twoside]{article}
\usepackage[headings]{espcrc2}

\usepackage{graphicx}
\usepackage[figuresright]{rotating}

\newcommand{\seff}{\sin^2\theta^{\mbox{\footnotesize
      lept}}_{\mbox{\footnotesize eff}}}

\title{
{\normalsize \tt WUE-ITP-2006-002\\}
\vspace*{.5cm}
Bosonic corrections to the effective leptonic weak mixing angle
  at the two-loop level}

\author{M. Czakon\address{Institut f\"ur Theoretische Physik und
    Astrophysik, Universit\"at W\"urzburg, \\ Am Hubland, D-97074
    W\"urzburg, Germany \\
    \vspace{.2cm}
    Institute of Physics, University of Silesia, ul. Uniwersytecka
    4, PL-40007 Katowice, Poland}
    \thanks{The work of M.C. was supported by the Sofja
      Kovalevskaja Award of the Alexander von Humboldt Foundation
      sponsored by the German Federal Ministry of Education and
      Research, and by the Polish State Committee for Scientific
      Research (KBN) for the research project in years
      2004-2005. The work of M.A. is supported by  BMBF grant No
      04-160.}
    M. Awramik\address{II. Institut f\"ur Theoretische Physik,
      Universit\"at Hamburg, \\ Luruper Chaussee 149, D-22761 Hamburg,
      Germany \\
      \vspace{.2cm}
      Institute of Nuclear Physics PAS, Radzikowskiego 152, PL-31342
      Cracow, Poland}
    A. Freitas\address{Institut f\"ur Theoretische Physik,
      Universit\"at Z\"urich, \\ Winterthurerstrasse 190, CH-8057
      Z\"urich, Switzerland}}
       
\begin{document}

\begin{abstract}
  Details of the recent calculation of the two-loop bosonic corrections
  to the effective leptonic weak mixing angle are presented. In
  particular, the expansion in the difference of the W and Z boson
  masses is studied and some of the master integrals needed are given in
  analytic form.
\end{abstract}

\maketitle

\section{INTRODUCTION}

The effective leptonic weak mixing angle, $\seff$ , is a crucial
observable in the indirect determination of the Higgs boson mass from
precision experiments up to the Z boson mass scale. Defined through the
vertex form factors of the $Z$ at its mass shell
\begin{equation}
\seff = \frac{1}{4} \left( 1-\Re \left( \frac{g_V(M_Z^2)}{g_A(M_Z^2)}
\right) \right), 
\end{equation}
it is an UV and IR finite quantity. The current experimental precision
requires a careful analysis of the error on the theory side. Here,
several contributions have been calculated over the years. In
particular, two- and three-loop QCD \cite{qcd2l,qcd3l}, two-loop electroweak
fermionic \cite{fermionic1,fermionic2}, and parts of the three-loop
electroweak contributions \cite{electroweak3l} are known. Recently,
also the Higgs boson mass dependence of the 
purely bosonic electroweak graphs has been obtained \cite{mhdependence}. The
computation of the $W$ boson mass from $\mu$ decay,
which is part of the final prediction for $\seff$ is more
complete, since the two-loop electroweak 
corrections have been fully evaluated \cite{muon1,muon2}.

It is important to note,
that the prediction from \cite{fermionic1} has been used by the LEP
Electroweak Working Group for their final report \cite{lepfinal}. It
is, therefore, necessary, to make sure that the theory error is not
underestimated due to the lacking bosonic electroweak corrections. To
this end, we have calculated the appropriate diagrams \cite{bosonic}
and shown that they fall well within our estimates. It is the purpose
of this work to present the details of our computation. 

\section{MASS DIFFERENCE EXPANSION}

The most complicated bosonic diagrams needed for $\seff$ are
two-loop vertices with up to three different masses,
$M_W$ , $M_Z$ and $M_H$. It is only natural \cite{massdifference} to
exploit the small difference between $M_W$ and $M_Z$ to reduce the
number of scales. Furthermore, since the authors of
\cite{mhdependence} have evaluated the Higgs boson mass 
dependence of the result normalized at $M_H$ = 100 GeV, it is only
necessary to evaluate the graphs at the same point at first (of course
an independent check of the mass dependence is also required). This
allows a further mass difference expansion, 
with expansion parameter $s^2_H = 1 - M^2_H/M^2_Z$ .

It is clear that if there are thresholds at $M_W = M_Z$, then starting
from some order in the expansion in $s^2_W = 1 - M^2_W /M^2_Z$, we are
going to encounter divergences. A 
suitable technique to recover the correct value is the expansion by
regions method \cite{smirnov}. The idea is to analyze the momentum
regions, which can contribute to the integral,and expand the integrand
in each region suitably performing the integration in dimensional
regularization. In the case, when only one line can go on-threshold,
there are just two regions: ultrasoft, (us), where
$k_{1,2} \sim s^2_W M_Z$; and hard, (h), where $k_{1,2} \sim M_Z$. The
classification of the size of the momenta has to be supplied by a
momentum routing, which clearly needs to be such that the selected line
goes on-threshold when $s_W = 0$.

\begin{figure}
  \includegraphics[width=8cm]{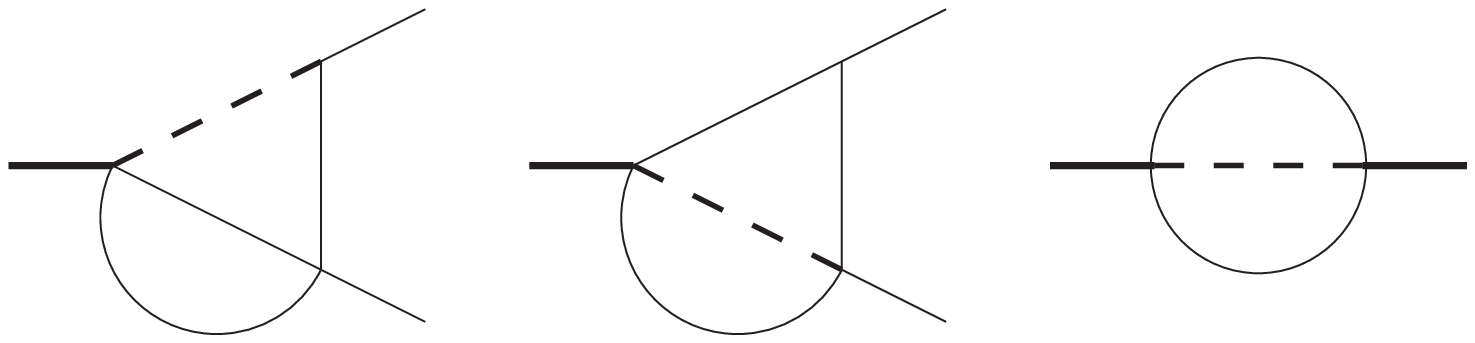}
  $I_1$ \hspace{2.3cm} $I_2$ \hspace{2.3cm} $I_3$
  \caption{\label{threshold} Ultrasoft graphs in the mass
    difference expansion. Dashed lines go on-threshold, when $M_W =
    M_Z$, and thin solid lines are massless.}
\end{figure}

In the (us) region, one encounters three possible topologies, 
Fig.~\ref{threshold}. The expansion transforms a Feynman integral into
an unusual form, since
\begin{equation}
  \label{eq2}
  \frac{1}{(k_{1,2}-p)^2-M_W^2} \sim \frac{1}{-2k_{1,2}p+s_W^2 M_Z^2},
\end{equation}
and
\begin{equation}
  \frac{1}{(k_{1,2}-p_{1,2})^2} \sim \frac{1}{-2k_{1,2}p_{1,2}}.
\end{equation}
The last integral in Fig.~\ref{threshold}, $I_3$, with the above
substitutions 
(actually just Eq.~(\ref{eq2})) has been evaluated in
\cite{smirnov}. The other two turn out to be reducible by the
integration-by-parts technique to the last case. In practice this
exercise has been left to {\it IdSolver} \cite{idsolver}, with the result
\begin{equation}
I_1 = \frac{1}{s_W^2 M_Z^2} \frac{4\epsilon - 3}{1-2\epsilon} I_3,
\end{equation}
and
\begin{equation}
I_2 = \frac{1}{s_W^2 M_Z^2}\left(\frac{3}{\epsilon}-4\right) I_3,
\end{equation}
where
\begin{eqnarray}
I_3 &=& \left( i\pi^{d/2}e^{-\epsilon \gamma_E}\right)^2
\Gamma^2(1-\epsilon) \Gamma(4\epsilon-3) \\
&\times& (M_Z^2)^{1-2\epsilon}(-s_W^2)^{3-4\epsilon}. \nonumber
\end{eqnarray}
Since there is no problem with the evaluation of the master integrals,
the rest of the calculation has been reduced to the application of
{\it IdSolver} to all integrals with numerators and dots.

\section{MASTER INTEGRALS}

The integrals in the hard momentum region,(h), have the usual Feynman
integrand structure, with some propagators on-threshold. In
practice this means that each 
one of them is given by a series in $\epsilon$ with numeric coefficients
multiplied by a trivial scaling factor. The problem is now that we have
spurious poles in the reduction to master integrals and we need
analytic results for constant parts of some nontrivial integrals in
order to check the exact cancellation of divergences. Of course, one
could think of constructing an $\epsilon$-finite basis in the sense of
\cite{basis}, but we would in the end have to evaluate integrals with
collinear singularities (even though just to finite parts). Therefore
we tried to find a basis that would produce a relatively small number of
spurious poles, but in front of simple integrals.

\begin{figure}
  \includegraphics[width=7cm]{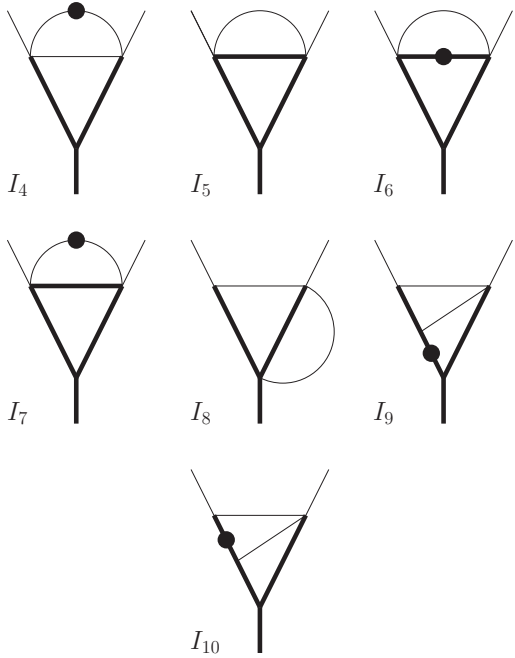}
  \caption{\label{hard} Some hard momentum region graphs needed to
    ${\cal O}(\epsilon)$. Thick lines are massive and on-shell,
    whereas thin are massless. A dot represents a squared propagator.}
\end{figure}

Indeed, there are 73 master integrals, 1 needed to ${\cal O}(\epsilon^3)$, 6 to
${\cal O}(\epsilon^2)$, and 26 to ${\cal O}(\epsilon)$. Most of
the integrals needed to high orders in the 
$\epsilon$-expansion could be found in the literature. In particular,
several vertex integrals have been given in \cite{vertices}. However, 7
integrals remained to be evaluated analytically down to the finite part,
Fig.~\ref{hard}. To this end, we used series
representations in the small external momentum regime, which were
subsequently improved with the help of conformally mapped Pad\'e
approximants, and resummed empirically with the PSLQ algorithm. The
expansions were derived from Mellin-Barnes representations and
analyzed with the help of the MB package \cite{mb}. With the usual
$\overline{\mbox{MS}}$ normalization of the momentum integrations,
{\it i.e.} $e^{\epsilon \gamma_E}/i\pi^{d/2} \int d^d k$
per loop, the results are
\begin{eqnarray}
I_4 &=& \frac{\pi^2}{9\epsilon}+\frac{3\pi
  \mbox{S2}}{\sqrt{3}}-\frac{2}{9}\zeta_3, \\
I_5 &=& \frac{1}{2\epsilon^2} + \frac{1}{\epsilon} \left(\frac{5}{2} -
  \frac{\pi}{\sqrt{3}} \right) +\frac{19}{2} + \frac{\pi^2}{18}
  \nonumber \\ &-&
  \frac{9 \mbox{S2}}{4} - \frac{5\pi}{\sqrt{3}} + \frac{\pi \log
  3}{\sqrt{3}} + \frac{9\pi \mbox{S2}}{2\sqrt{3}} - \frac{8}{3}
  \zeta_3, \nonumber \\
I_6 &=& \frac{9\pi \mbox{S2}}{2\sqrt{3}} - \frac{8}{3} \zeta_3,
  \nonumber \\
I_7 &=& \frac{\pi^2}{18\epsilon} - \frac{15 \pi \mbox{S2}}{2\sqrt{3}}
  + \frac{23}{9} \zeta_3, \nonumber \\
I_8 &=& \frac{1}{2\epsilon^2} + \frac{3}{2\epsilon} + \frac{5}{2} +
  \frac{\pi^2}{36} + \frac{\pi}{\sqrt{3}} -
  \frac{9\pi\mbox{S2}}{2\sqrt{3}} + \frac{1}{3} \zeta_3, \nonumber \\
I_9 &=& -\frac{\pi^3}{54\sqrt{3}} + \frac{3\pi \mbox{S2}}{2\sqrt{3}} +
  \frac{2}{9} \zeta_3, \nonumber \\
I_{10} &=& \frac{3\pi \mbox{S2}}{\sqrt{3}} - \frac{5}{9} \zeta_3, \nonumber
\end{eqnarray}
where S2 = $4/(9\sqrt{3})\mbox{Cl}_2(\pi/3)$, with $\mbox{Cl}_2(x) =
\Im(\mbox{Li}_2(e^{ix}))$, and the mass has been set to unity.The
${\cal O}(\epsilon)$ parts of these integrals and finite parts of the
remaining ones have been evaluated numerically, either from small momentum
expansions,or from integral representations. There was no need to
invest into an analytic result.

\begin{figure}
  \includegraphics[width=8cm]{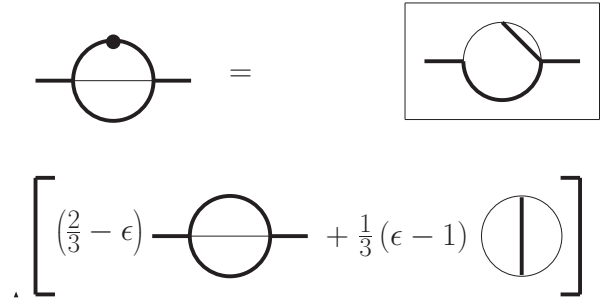}
  \caption{\label{rel} Exact relation between 3-line
    masters on-shell. The integral, of which IBPs contain this relation
    is shown in the frame.}
\end{figure}

During the reduction of the integrals
to masters, and in particular of a four line propagator, an
interesting relation has been discovered, Fig.~\ref{rel}. Such a relation
could certainly not be found during the reduction of the three line
propagator graph, since the vacuum graph has the same number of lines,
but a different mass and momentum distribution. In fact, reducing
the propagator leads to the two masters, the dotted and the
undotted. This result demonstrates that the integration-by-parts
relations do not exhaust all the rational function coefficient linear
relations between integrals with a fixed upper number of lines in general.

Note, that the fact that there is a simple relation between the two
3-line propagator master integrals has been known since \cite{kalmykov}, but
there,it was not identified as a relation between three integrals.

\section{VERTEX CONTRIBUTIONS}

Summing up all of the vertex contributions in Feynman gauge, with the
mass parameter of dimensional regularization set to $M_Z$, and $M_H$ =
100 GeV, one obtains
\begin{eqnarray}
\label{sum}
&& \!\!\!\!\!-4.4 \times 10^{-6} \frac{1}{s^2_W} ( 1 - 1.1s^2_W - 3.9s^4_W -
  0.8s^6_W \nonumber \\
&& +0.2s^8_W + 1.1 \times 10^{-3} \log(s^2_W )s^8_W + ...)
\end{eqnarray}
Even though just
part of the calculation, this formula is interesting by itself. First
of all, the remaining part of the result can be evaluated to very high
precision for any masses, since it is expressed by at most two-loop
propagators, and therefore, the above series determines the final
precision. Second, Eq.~(\ref{sum}) proves that one should resist the temptation
to approximate the final contribution by the leading term, which is
relatively easily evaluated by setting
$M_W = M_Z$ in all lines. In fact, because $s^2_W \simeq 1/4$, the sum
of the first three terms is equal to half of the leading one. Finally,
the ultrasoft contribution connected to the logarithm of $s_W$ starts very
late in the expansion and has a small coefficient, which means that
one could have obtained a reliable approximation without the strategy
of regions. 

\begin{figure}
  \includegraphics[width=8cm]{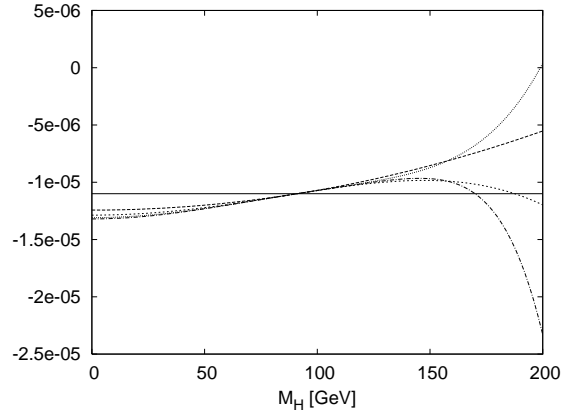}
  \caption{\label{higgs} Behavior of the subsequent terms of the
    expansion in $s_H^2$ of the sum of the vertex graphs.}
\end{figure}

The convergence of the expansion can also be studied by looking at the
dependence on $M_H$.The first five terms of the expansion are
\begin{eqnarray}
\!\!\!\!\!-1.1 \times 10^{-5} ( 1 + 0.13s^2_H + 0.04s^4_H + 0.02s^6_H +
   0.01s^8_H), \nonumber
\end{eqnarray}
which shows nice convergence at $M_H = 0$, corresponding to
$s^2_H = 1$. At the other end of the scale we obtain an alternating series,
which strongly diverges even below the threshold for Higgs boson decay
into a pair of gauge bosons, as shown in Fig.~\ref{higgs}. It turns out
that this behavior can be of advantage. In particular, being alternating this
series should lend itself nicely to Pad\'e resummation techniques,
closing the gap between the mass difference and large Higgs boson
mass expansion. A similar phenomenon has been observed during the
computation of the bosonic corrections to $\mu$ decay \cite{muon2}.

\section{SUMMARY}

We have discussed computational techniques used in the recent
computation of the complete bosonic contributions to
the effective leptonic weak mixing angle. As a by product, we have
presented new results for on-shell two-loop vertices and a relation
between on-shell propagator master integrals.

\section{ACKNOWLEDGMENT}

M.C. would like to thank M. Yu. Kalmykov for drawing his attention to
\cite{kalmykov}.


\begin{thebibliography}{99}

\bibitem{qcd2l}
A.~Djouadi and C.~Verzegnassi,
%``Virtual Very Heavy Top Effects In Lep / Slc Precision Measurements,''
Phys.\ Lett.\ B {\bf 195} (1987) 265;
%%CITATION = PHLTA,B195,265;%%
A.~Djouadi,
%``O (Alpha Alpha-S) Vacuum Polarization Functions Of The Standard Model Gauge
%Bosons,''
Nuovo Cim.\ A {\bf 100} (1988) 357;
%%CITATION = NUCIA,A100,357;%%
B.~A.~Kniehl,
%``Two Loop Corrections To The Vacuum Polarizations In Perturbative QCD,''
Nucl.\ Phys.\ B {\bf 347} (1990) 86;
%%CITATION = NUPHA,B347,86;%%
F.~Halzen and B.~A.~Kniehl,
%``Delta R Beyond One Loop,''
Nucl.\ Phys.\ B {\bf 353} (1991) 567;
%%CITATION = NUPHA,B353,567;%%
B.~A.~Kniehl and A.~Sirlin,
%``Dispersion relations for vacuum polarization functions in electroweak
%physics,''
Nucl.\ Phys.\ B {\bf 371} (1992) 141;
%%CITATION = NUPHA,B371,141;%%
%B.~A.~Kniehl and A.~Sirlin,
%``On the effect of the t anti-t threshold on electroweak parameters,''
Phys.\ Rev.\ D {\bf 47} (1993) 883;
%%CITATION = PHRVA,D47,883;%%
A.~Djouadi and P.~Gambino,
%``Electroweak gauge bosons selfenergies: Complete QCD corrections,''
Phys.\ Rev.\ D {\bf 49} (1994) 3499
[Erratum-ibid.\ D {\bf 53} (1996) 4111].
%%CITATION = HEP-PH 9309298;%%

\bibitem{qcd3l}
K.~G.~Chetyrkin, J.~H.~K\"uhn and M.~Steinhauser,
%``QCD corrections from top quark to relations between electroweak parameters to order alpha-s**2,''
Phys.\ Rev.\ Lett.\  {\bf 75} (1995) 3394;
%%CITATION = HEP-PH 9504413;%%
%K.~G.~Chetyrkin, J.~H.~Kuhn and M.~Steinhauser,
%``Three-loop polarization function and O(alpha(s)**2) corrections to the
%production of heavy quarks,''
Nucl.\ Phys.\ B {\bf 482} (1996) 213.
%%CITATION = HEP-PH 9606230;%%

\bibitem{fermionic1}
M.~Awramik, M.~Czakon, A.~Freitas and G.~Weiglein,
%``Complete two-loop electroweak fermionic corrections to
%sin**2(Theta(lept)(eff)) and indirect determination of the Higgs
%boson  mass,'' 
Phys.\ Rev.\ Lett.\  {\bf 93} (2004) 201805.
%[arXiv:hep-ph/0407317].
%%CITATION = HEP-PH 0407317;%%

\bibitem{fermionic2}
W.~Hollik, U.~Meier and S.~Uccirati,
%``The effective electroweak mixing angle sin**2(theta(eff)) with two-loop
%fermionic contributions,''
Nucl.\ Phys.\ B {\bf 731} (2005) 213.
%[arXiv:hep-ph/0507158].
%%CITATION = HEP-PH 0507158;%%

\bibitem{electroweak3l}
M.~Faisst, J.~H.~Kuhn, T.~Seidensticker and O.~Veretin,
%``Three loop top quark contributions to the rho parameter,''
Nucl.\ Phys.\ B {\bf 665} (2003) 649;
%[arXiv:hep-ph/0302275].
%%CITATION = HEP-PH 0302275;%%
R.~Boughezal, J.~B.~Tausk and J.~J.~van der Bij,
%``Three-loop electroweak correction to the rho parameter in the large  Higgs
%mass limit,''
Nucl.\ Phys.\ B {\bf 713} (2005) 278;
%[arXiv:hep-ph/0410216].
%%CITATION = HEP-PH 0410216;%%
R.~Boughezal, J.~B.~Tausk and J.~J.~van der Bij,
%``Three-loop electroweak corrections to the W-boson mass and
%sin**2(theta(lept)(eff)) in the large Higgs mass limit,''
Nucl.\ Phys.\ B {\bf 725} (2005) 3.
%[arXiv:hep-ph/0504092].
%%CITATION = HEP-PH 0504092;%%

\bibitem{mhdependence}
W.~Hollik, U.~Meier and S.~Uccirati,
%``Higgs-mass dependence of the effective electroweak mixing angle
%sin**2(theta(eff)) at the two-loop level,''
Phys.\ Lett.\ B {\bf 632} (2006) 680.
%[arXiv:hep-ph/0509302].
%%CITATION = HEP-PH 0509302;%%

\bibitem{muon1}
A.~Freitas, W.~Hollik, W.~Walter and G.~Weiglein,
%``Complete fermionic two-loop results for the M(W)-M(Z)
%interdependence,''
Phys.\ Lett.\ B {\bf 495} (2000) 338
[Erratum-ibid.\ B {\bf 570} (2003) 260],
%%CITATION = HEP-PH 0007091;%%
%A.~Freitas, W.~Hollik, W.~Walter and G.~Weiglein,
%``Electroweak two-loop corrections to the M(W) - M(Z) mass correlation
%in  the standard model,''
Nucl.\ Phys.\ B {\bf 632} (2002) 189
[Erratum-ibid.\ B {\bf 666} (2003) 305];
%%CITATION = HEP-PH 0202131;%%
M.~Awramik and M.~Czakon,
%``Complete two loop electroweak contributions to the muon lifetime in
%the  standard model,''
Phys.\ Lett.\ B {\bf 568} (2003) 48.
%%CITATION = HEP-PH 0305248;%%

\bibitem{muon2}
M.~Awramik and M.~Czakon,
%``Complete two loop bosonic contributions to the muon lifetime in the
%standard model,''
Phys.\ Rev.\ Lett.\  {\bf 89} (2002) 241801;
%%CITATION = HEP-PH 0208113;%%
A.~Onishchenko and O.~Veretin,
%``Two-loop bosonic electroweak corrections to the muon lifetime and
%M(Z) M(W) interdependence,''
Phys.\ Lett.\ B {\bf 551} (2003) 111;
%%CITATION = HEP-PH 0209010;%%
M.~Awramik, M.~Czakon, A.~Onishchenko and O.~Veretin,
%``Bosonic corrections to Delta(r) at the two loop level,''
Phys.\ Rev.\ D {\bf 68} (2003) 053004.
%%CITATION = HEP-PH 0209084;%%

\bibitem{lepfinal}
the LEP Collaborations [OPAL Collaboration],
%``A combination of preliminary electroweak measurements and constraints on the
%standard model,''
arXiv:hep-ex/0511027.
%%CITATION = HEP-EX 0511027;%%

\bibitem{bosonic}
M.~Awramik, M.~Czakon, A.~Freitas, {\it in preparation}.

\bibitem{massdifference}
F.~Jegerlehner, M.~Y.~Kalmykov and O.~Veretin,
%``MS-bar vs. pole masses of gauge bosons: Electroweak bosonic two-loop
%corrections,''
Nucl.\ Phys.\ B {\bf 641} (2002) 285;
%[arXiv:hep-ph/0105304].
%%CITATION = HEP-PH 0105304;%%
F.~Jegerlehner, M.~Y.~Kalmykov and O.~Veretin,
%``MS-bar vs pole masses of gauge bosons. II: Two-loop electroweak fermion
%corrections,''
Nucl.\ Phys.\ B {\bf 658} (2003) 49.
%[arXiv:hep-ph/0212319].
%%CITATION = HEP-PH 0212319;%%

\bibitem{smirnov}
V.~A.~Smirnov, {\it ``Applied asymptotic expansions in momenta and
masses''}, Berlin, Germany, Springer (2002).

\bibitem{idsolver}
M.~Czakon, {\it ``DiaGen/IdSolver''}, {\it unpublished}; {\it see also}
M.~Awramik, M.~Czakon, A.~Freitas and G.~Weiglein,
%``Two-loop fermionic electroweak corrections to the effective leptonic  weak
%mixing angle in the standard model,''
Nucl.\ Phys.\ Proc.\ Suppl.\  {\bf 135} (2004) 119.
%[arXiv:hep-ph/0408207].
%%CITATION = HEP-PH 0408207;%%

\bibitem{basis}
K.~G.~Chetyrkin, M.~Faisst, C.~Sturm and M.~Tentyukov,
%``e-finite basis of master integrals for the integration-by-parts method,''
arXiv:hep-ph/0601165.
%%CITATION = HEP-PH 0601165;%%

\bibitem{vertices}
U.~Aglietti and R.~Bonciani,
%``Master integrals with one massive propagator for the two-loop  electroweak
%form factor,''
Nucl.\ Phys.\ B {\bf 668} (2003) 3;
%[arXiv:hep-ph/0304028].
%%CITATION = HEP-PH 0304028;%%
U.~Aglietti and R.~Bonciani,
%``Master integrals with 2 and 3 massive propagators for the 2-loop
%electroweak 
%form factor: Planar case,''
Nucl.\ Phys.\ B {\bf 698} (2004) 277.
%[arXiv:hep-ph/0401193].
%%CITATION = HEP-PH 0401193;%%

\bibitem{mb}
M.~Czakon,
%``Automatized analytic continuation of Mellin-Barnes integrals,''
arXiv:hep-ph/0511200.
%%CITATION = HEP-PH 0511200;%%

\bibitem{kalmykov}
A.~I.~Davydychev and M.~Y.~Kalmykov,
%``New results for the epsilon-expansion of certain one-, two- and  three-loop
%Feynman diagrams,''
Nucl.\ Phys.\ B {\bf 605} (2001) 266.
%[arXiv:hep-th/0012189].
%%CITATION = HEP-TH 0012189;%%

\end{thebibliography}
\end{document}